\documentclass[prb,preprintnumbers,tightenlines,12pt,showpacs,nobibnotes,byrevtex,amssymb,altaffilletter]{revtex4}

\newcommand{\ket}[1]{|#1\rangle}
\newcommand{\bra}[1]{\langle #1 |}
\newcommand{\mymid}{\,:\,}

\newtheorem{proposition}{Proposition}
\newtheorem{theorem}[proposition]{Theorem}
\newtheorem{lemma}[proposition]{Lemma}
\newtheorem{definition}[proposition]{Definition}
\newtheorem{remark}[proposition]{Remark}
\newtheorem{corollary}[proposition]{Corollary}

\begin{document}
\preprint{quant-ph/0105151}
\title{Lower bound for the quantum capacity of
a discrete memoryless quantum channel}
\thanks{To appear in J. Math.\ Phys.\ Special Issue on Quantum Information Theory, Sept.\ 2002.}
\author{Ryutaroh Matsumoto}
\email{ryutaroh@rmatsumoto.org}
\homepage{http://www.rmatsumoto.org}
\author{Tomohiko Uyematsu}
\email{uyematsu@ieee.org}
\homepage{http://www.it.ss.titech.ac.jp}
\affiliation{Department of Communications and Integrated Systems,
Tokyo Institute of Technology, 152-8552 Japan}
\begin{abstract}
We generalize the random coding argument of stabilizer codes
and derive a lower bound on the quantum capacity
of an arbitrary discrete memoryless quantum channel.
For the depolarizing channel,
our lower bound coincides with that obtained
by Bennett et~al.
We also slightly improve
the quantum Gilbert-Varshamov bound
for general stabilizer codes, and
establish an analogue of the quantum Gilbert-Varshamov bound
for linear stabilizer codes.
Our proof is restricted to the binary quantum channels,
but its extension of to $l$-adic channels
is straightforward.
\end{abstract}
\maketitle
\section{Introduction}
The quantum capacity of a quantum channel
is the amount of quantum states that can be reliably transmitted
through the channel.
It is one of fundamental
unsolved problems in the quantum information theory.
Except the quantum erasure channel, we know only lower and upper bounds
for the quantum capacity
of a quantum channel,
and in addition,
a tight lower bound is not known for a general memoryless
quantum channel.
In this paper we shall demonstrate a lower bound on the capacity
of a general memoryless quantum channel.
A quantum channel is said to be memoryless
if the state change of one
transmitted quantum system (of the fixed degree of freedom)
is statistically independent of the state change of another.

The problem of quantum capacity has attracted great attention,
and rapid progress has been made.
To be precise,
the quantum capacity of a binary memoryless channel
$\Gamma$ is the maximum number $Q(\Gamma)$
such that for any rate $R< Q(\Gamma)$
and any $\epsilon > 0$
there exists an $[[n,k]]$ quantum code $Q$
with $k/n \geq R$ such that
the fidelity between the recovered state and the original
state $\ket{\varphi}\in Q$ is at least $1-\epsilon$ for any
$\ket{\varphi}$.\cite{bennett97,bennett96}
In Refs.~\onlinecite{bennett97,bennett96},
the authors obtained
the exact capacity of the quantum erasure channel,
and showed lower and upper bounds for that of the quantum depolarizing channel.
The same lower bounds for those channels were also
obtained in Ref.~\onlinecite{gottesmanthesis}
by using random coding of the stabilizer codes
introduced in Refs.~\onlinecite{calderbank97,calderbank98,gottesman96}.
After that, DiVincenzo et~al.\cite{divincenzo98}
improved the lower bound for a depolarizing channel
by using nonrandom stabilizer codes.
The upper bound of the depolarizing channel
was improved in Refs.~\onlinecite{bruss98,rains98,vedral98},
and generalized to asymmetric depolarizing channels in
Ref.~\onlinecite{cerf00}.
An apparently different definition of the quantum capacity
was formalized in Ref.~\onlinecite{barnum98},
in which an upper bound of a general memoryless quantum channel
was established by
using the notion of coherent information introduced
in Ref.~\onlinecite{schumacher96b}.
It is informally argued in Ref.~\onlinecite{lloyd97}
that the upper bound in Ref.~\onlinecite{barnum98}
is achieved by random coding over a general memoryless channel.
Barnum et~al.\cite{barnum00} showed that
the definitions of quantum capacity in Refs.~\onlinecite{barnum98,bennett97,bennett96}
were equivalent.

It is the random coding that the most commonly used technique in
classical information theory to show that a specific rate is
achieved by a code in a specific class of codes.
For example, Elias showed that the capacity of the
binary symmetric channel is achieved by binary linear codes
using random coding\cite{elias55} (a readable proof of this fact
can be found in Sec.~6.2 of Ref.~\onlinecite{gallager68}).
A proof by random coding is usually proceeded as follows:
one first calculates the average of error probability of
all codes in a specific class of codes of the same rate
and the same code length,
then shows that the average converges to $0$ as the code length
increases, and finally concludes that there exists
at least one sequence of codes of the fixed rate
with which the error probability converges to $0$.

The technique of random coding is also used in the
quantum information theory.
Gottesman\cite{gottesmanthesis} showed using random coding
that the lower bound on the quantum capacity
of the depolarizing channel\cite{bennett96}
can be achieved by stabilizer codes.
However, his proof does not seem to extend easily
to the case of general memoryless channel.
The aim of the present paper is to derive
a lower bound on the quantum capacity of a general
memoryless channel by using random coding of
stabilizer codes.
In our argument of random coding,
we shall use the fidelity\cite{jozsa94,uhlmann76}
as a replacement
of error probability in the classical random coding,
and use the idea behind
the proof of the quantum Gilbert-Varshamov
bound for the stabilizer codes.\cite{calderbank97}
As a byproduct,
we also improve the quantum Gilbert-Varshamov bound
for stabilizer codes.
Our improved bound (Remark \ref{remark5}) is slightly better than
the quantum Gilbert-Varshamov bound
for general codes.\cite{ekert96}

As a natural consequence of
the quantum Gilbert-Varshamov bound
\cite{calderbank97,ekert96} and the fidelity bound
of $t$-error correcting quantum codes,\cite{knill97,matsumotoerror,preskill98}
we can also
derive lower bounds for the quantum capacity of a general memoryless
quantum channel.
However, for the depolarizing channel,
the derived lower bounds are much smaller than
that obtained in Ref.~\onlinecite{bennett96}.
In contrast to this,
our lower bound coincides with the bound
in Ref.~\onlinecite{bennett96} for the depolarizing channel.

It is interesting whether the proposed lower bound
is achieved by a subclass of general stabilizer codes.
We also show that the random coding of linear stabilizer
codes yields the same lower bound on the quantum capacity.
As a byproduct we obtain an analogue of the quantum Gilbert-Varshamov
bound for linear stabilizer codes (Remark \ref{gvlinear}),
which is asymptotically the same as that
for general quantum codes.\cite{ekert96}

The quantum channel considered in this paper
is discrete in the sense that
the channel carries finite-dimensional quantum systems,
and we do not touch the quantum capacity of a continuous
quantum channel recently studied in Refs.~\onlinecite{gottesman01,harrington01}.
Our proof is restricted to the binary quantum channels
for the simplicity of presentation,
but its extension to $l$-adic channels
is straightforward for prime $l$.

This paper is organized as follows:
In Sec.~\ref{sec2} we introduce notations and review
relevant research results.
In Sec.~\ref{sec3}
we derive a lower bound [Eq.~(\ref{condr2a})] for the quantum capacity
of an arbitrary discrete memoryless quantum channel
by random coding of stabilizer codes.

\section{Notations and preliminaries}\label{sec2}
In this section
we fix notations used in this paper,
and review known research results that are necessary
to establish our results.

\subsection{Quantum channel and its quantum capacity}
For a finite-dimensional complex Hilbert space $\mathcal{H}$,
let $\mathcal{S}(\mathcal{H})$ be the set of density operators
on $\mathcal{H}$, and $\mathcal{L}(\mathcal{H})$
the set of linear operators on $\mathcal{H}$.
The standard description of a quantum channel
is the completely positive trace-preserving map (CP map).\cite{holevo74,holevo77,schumacher96}
Suppose that we send a state $\rho \in \mathcal{S}(\mathcal{H})$.
The statistical ensemble of the received states
is described as $\Gamma(\rho)$ by a CP map $\Gamma$.

Suppose that we send a state $\rho \in \mathcal{S}
(\mathcal{H}^{\otimes n})$
through a quantum channel.
The quantum channel is said to be \emph{memoryless}
if the received state is described as
$\Gamma^{\otimes n}(\rho)$ for all $\rho \in \mathcal{S}(
\mathcal{H}^{\otimes n})$
and for some CP map $\Gamma$ on $\mathcal{L}(\mathcal{H})$.

Fidelity is a measure of closeness
between two quantum states.
The fidelity $F$ between a pure state $\ket{\varphi}\in \mathcal{H}$
and a state $\rho \in \mathcal{S}(\mathcal{H})$
is defined by $\bra{\varphi} \rho \ket{\varphi}$.\cite{jozsa94,uhlmann76}
We have $0\leq F \leq 1$ and
two states are closer if the fidelity between them is larger.

Let $H_2$ be the two-dimensional complex Hilbert space.
Unless otherwise stated we consider the binary memoryless quantum channel,
that is,
when we send $\rho \in \mathcal{S}(H_2^{\otimes n})$
we receive $\Gamma^{\otimes n}(\rho)$,
where $\Gamma$ is a CP map on $\mathcal{L}(H_2)$.
We shall identify a binary memoryless channel
with a CP map on $\mathcal{L}(H_2)$.

A binary $[[n,k]]$ quantum code $Q$
is a $2^k$-dimensional subspace of $H_2^{\otimes n}$.
The rate of an $[[n,k]]$ quantum code
is $k/n$.
The quantum capacity of a binary memoryless channel
$\Gamma$ is the maximum number $Q(\Gamma)$
such that for any rate $R< Q(\Gamma)$
and any $\epsilon>0$
there exists an $[[n,k]]$ quantum code $Q$
with $k/n \geq R$ such that
the fidelity between the recovered state and the original
state $\ket{\varphi}\in Q$ is at least $1-\epsilon$ for any
$\ket{\varphi}$.\cite{bennett97,bennett96}

\subsection{Fidelity bound of the quantum error correction}
In this subsection
we review Preskill's lower bound on the fidelity
of quantum error correction in terms of
the set of uncorrectable errors of a quantum code.
Let
\[
\sigma_x = \left(\begin{array}{cc}0&1\\ 1&0\end{array}\right),\;
\sigma_z = \left(\begin{array}{cc}1&0\\ 0&-1\end{array}\right),
\]
and $\mathcal{E} = \{ w_1 \otimes $ $\cdots$ $\otimes w_n\}$,
where $w_i$ is either $I$, $\sigma_x$, $\sigma_z$ or $\sigma_x\sigma_z$.
For a quantum code $Q$ and a fixed error correction process for $Q$,
an operator $M \in \mathcal{E}$ is said to be correctable if
the error correction process of $Q$ recovers
$M\ket{\varphi}$ to $\ket{\varphi}$ for all $\ket{\varphi}\in Q$.
An operator $M$ is uncorrectable if it is not correctable.
Let $\mathcal{E}_\mathrm{unc}\subset \mathcal{E}$
be the set of uncorrectable errors of a quantum code
$Q \subset H_2^{\otimes n}$.
Suppose that we send a pure state
$\ket{\varphi} \in Q$ through a binary memoryless channel
described by a CP map $\Gamma$ on $\mathcal{L}(H_2)$.
By a unitary representation of a CP map,\cite{book:kraus}
there exists a finite-dimensional Hilbert space
$H_\mathrm{env}$, a pure state $\ket{0_\mathrm{env}} \in H_\mathrm{env}$
and a unitary operator $U$ on $H_2^{\otimes n} \otimes H_\mathrm{env}$
such that
\begin{equation}
\Gamma(\rho) = \mathrm{Tr}_{H_\mathrm{env}}
(U (\rho \otimes \ket{0_\mathrm{env}}\bra{0_\mathrm{env}})U^*)
\label{unirep}
\end{equation}
for all $\rho \in \mathcal{S}(H_2^{\otimes n})$,
where $\mathrm{Tr}_{H_\mathrm{env}}$ is the partial trace
over ${H_\mathrm{env}}$.
Since $\mathcal{E}$ is a basis of
$\mathcal{L}(H_2^{\otimes n})$ we can write $U$ in Eq.~(\ref{unirep})
as
\[
U = \sum_{M \in \mathcal{E}} M \otimes L_M,
\]
where $L_M$ is a linear operator on $H_\mathrm{env}$.
Preskill proved the following theorem in Sec.~7.4
of Ref.~\onlinecite{preskill98}.

\begin{theorem}\label{thm1}
Let $Q$ and  $\mathcal{E}_\mathrm{unc}$ be as above.
When we send a pure state $\ket{\varphi} \in Q$,
the fidelity between $\ket{\varphi}$ and the recovered state
is not less than
\[
1-
\left\|
\sum_{M \in \mathcal{E}_\mathrm{unc}} M\ket{\varphi}
\otimes L_M \ket{0_\mathrm{env}}\right\|^2,
\]
where $\|\cdot\|$ denotes the norm of a vector.
\end{theorem}

\subsection{Stabilizer codes and their error correction process}\label{sec:stabilizer}
In this subsection we review stabilizer quantum codes
introduced in Refs.~\onlinecite{calderbank97,calderbank98,gottesman96}.
Let $E=\{ \pm w_1 \otimes $ $\cdots$ $\otimes w_n\}$,
where $w_i$ is either $I$, $\sigma_x$, $\sigma_z$ or $\sigma_x\sigma_z$,
$S$ a commutative subgroup of $E$, and
\[
S' = \{ M\in E\mymid \forall N\in S,\; MN=NM\}.
\]
A stabilizer code $Q$ is defined as a simultaneous eigenspace
of all matrices in $S$.
If $S'$ has $2^{n+k+1}$ elements
then $\dim Q = 2^k$.
The set of simultaneous eigenspaces of $S$ is equal
to $\{ MQ \mymid M \in \mathcal{E}\}$,
where $MQ = \{ M\ket{\varphi}\mymid \ket{\varphi}\in Q\}$.

We shall describe the error correction process
of a stabilizer code.
Suppose that we send a pure state $\ket{\varphi}\in Q$
and received $\rho \in \mathcal{S}(H_2^{\otimes n})$.
We measure an observable of $H_2^{\otimes n}$
whose eigenspaces are the same as those of $S$.
Then the received state $\rho$ is projected to 
a state $\rho'$ that is an ensemble of pure states
in some eigenspace $Q'$ of $S$.
For $M = \pm w_1 \otimes $ $\cdots$ $\otimes w_n \in E$
we define the weight $w(M)$ of $M$
by $\sharp \{i \mymid w_i \neq I\}$,
where $\sharp$ denotes the number of elements in a set.
Let $M \in \mathcal{E}$ such that $MQ = Q'$ and that
if $MQ = M'Q$ for $M'\in\mathcal{E}$ then $w(M) \leq w(M')$.
We recover $\rho'$ to $M^{-1}\rho' (M^{-1})^*$.
With this error correction process
the set of uncorrectable errors is contained in
\begin{eqnarray}
&&\{ M\in \mathcal{E} \mymid
\mbox{there exists }
M'\in\mathcal{E} \mbox{ such that } w(M')\leq w(M),\nonumber\\
&&M'Q = MQ, \mbox{ and } MS \neq \pm M'S \}\nonumber\\
&=&\{ M\in \mathcal{E} \mymid
\mbox{there exists }
M'\in\mathcal{E} \mbox{ such that } w(M')\leq w(M),\nonumber\\
&&M'S' = MS', \mbox{ and } MS \neq \pm M'S \}.\label{uncorrectableset}
\end{eqnarray}

Hamada \cite{hamada01b} showed the following theorem based on
Theorem \ref{thm1}.
\begin{theorem}\label{thm2}
Notations as in Theorem \ref{thm1}.
Let $Q$ be a stabilizer code with the decoding process described above.
Then there exists a subspace $Q' \subset Q$ such that $\dim Q' =
\dim Q / 2$ and that
for all pure state $\ket{\varphi} \in Q'$,
the fidelity between $\ket{\varphi}$ and the recovered state is not less than
\begin{equation}
1-
2\sum_{M \in \mathcal{E}_\mathrm{unc}} 
\| L_M \ket{0_\mathrm{env}}\|^2.\label{preskillbound}
\end{equation}
\end{theorem}
Observe that the information rates of $Q$ and $Q'$ in Theorem \ref{thm2}
differ by $\log 2 / n$, which becomes negligible as $n \rightarrow
\infty$.
We call a subspace $Q'$ as a subcode of $Q$ as in the classical
coding theory.
We shall consider the subcode $Q'$ of a stabilizer code
$Q$ in the discussion of Sec.~\ref{sec3}.

\subsection{Symplectic geometry}
In this subsection we review the symplectic geometric interpretation
of stabilizer codes introduced in Refs.~\onlinecite{calderbank97,calderbank98}.
A symplectic geometry is a linear space with a nondegenerate
symplectic form.\cite{aschbacher00}
Let $\mathbf{F}_2$ be the finite field with $2$ elements.
For $\vec{a} = (a_1$, \ldots, $a_n) \in \mathbf{F}_2^n$ and
$\vec{b} = (b_1$, \ldots, $b_n) \in \mathbf{F}_2^n$,
we define $(\vec{a}|\vec{b})$ by
$(a_1$, \ldots, $a_n$, $b_1$, \ldots, $b_n)\in
\mathbf{F}_2^{2n}$ and
\[
f(\pm\sigma_x^{a_1}\sigma_z^{b_1}\otimes \cdots\otimes
\sigma_x^{a_n}\sigma_z^{b_n}) = (\vec{a}|\vec{b}).
\]
We also define the standard symplectic form of $(\vec{a}|\vec{b})$
and $(\vec{a}'|\vec{b}') \in \mathbf{F}_2^{2n}$ by
\begin{equation}
\langle \vec{a},\vec{b}'\rangle - \langle\vec{a}',\vec{b}\rangle,
\label{symplecticform}
\end{equation}
where $\langle \cdot,\cdot\rangle$ denotes the standard
inner product in $\mathbf{F}_2^n$.
For a subspace $C \subseteq \mathbf{F}_2^{2n}$
we denote
by $C^\perp$
the orthogonal space of $C$ with respect to (\ref{symplecticform}).
For a subgroup $S \subseteq E$,
$S$ is commutative if and only if
$f(S) \subseteq (f(S))^\perp$,
and $S' = f^{-1}((f(S))^\perp)$.

\subsection{Linear stabilizer codes and unitary geometry}
Calderbank et~al.\cite{calderbank98}
related stabilizer codes to classical
error-correcting codes and unitary geometry,
which is a linear space with a nondegenerate hermitian form.\cite{aschbacher00}
Let $\omega$ be a primitive element in $\mathbf{F}_4$,
and define
\[
g(\pm\sigma_x^{a_1}\sigma_z^{b_1}\otimes \cdots\otimes
\sigma_x^{a_n}\sigma_z^{b_n})
= \omega \vec{a} + \omega^2\vec{b} \in \mathbf{F}_4^n.
\]
For vectors $\vec{x}$, $\vec{y} \in \mathbf{F}_4^n$
we define an $\mathbf{F}_2$-bilinear map
\begin{equation}
\langle \vec{x}^2,\vec{y}\rangle - \langle \vec{x},
\vec{y}^2\rangle,\label{traceinner}
\end{equation}
where $\langle,\rangle$ denotes the standard inner product
in $\mathbf{F}_4^n$ and $\vec{x}^2 = (x_1^2$, \ldots, $x_n^2)$.
For an $\mathbf{F}_2$-linear subspace $C$ of $\mathbf{F}_4^n$,
let $C^\perp$ denotes the orthogonal space of $C$
with respect to (\ref{traceinner}).
For a subgroup $S \subseteq E$,
$S$ is commutative if and only if
$g(S) \subseteq (g(S))^\perp$,
and $S' = g^{-1}((g(S))^\perp)$.

For $\vec{x}$, $\vec{y}\in \mathbf{F}_4^n$,
we define the standard hermitian form of $\vec{x}$, $\vec{y}$ by
\begin{equation}
\tau(\vec{x},\vec{y}) = 
\langle \vec{x}^2, \vec{y}\rangle \label{hermitianinner},
\end{equation}
which is used only in Sec.~\ref{sec33}.
If $C$ is an $\mathbf{F}_4$-linear subspace of $\mathbf{F}_4^n$,
$C^\perp$ is equal to the orthogonal space of $C$
with respect to (\ref{hermitianinner}).
A stabilizer code constructed from an $\mathbf{F}_2$-linear
self-orthogonal space $C\subset\mathbf{F}_4^n$ is said to be
linear if $C$ is $\mathbf{F}_4$-linear.
This connection between binary stabilizer codes
and classical codes over $\mathbf{F}_4$ is generalized
to nonbinary case in Refs.~\onlinecite{ashikhmin00,bierbrauer00,matsumotouematsu00}.

\section{Lower bound on the quantum capacity}\label{sec3}
As described in the introduction,
we have to calculate the average of fidelity
over all the stabilizer codes, and show
that the average converges to $1$.
Strictly speaking, we shall use the subcode of a stabilizer code
introduced in Theorem \ref{thm2}.
This section is organized as follows:
In Sec.~\ref{sec30} we introduce a definition
of the distance of a quantum channel from the identity channel.
In Sec.~\ref{sec31}
we calculate the average of fidelity
over subcodes of general stabilizer codes of the fixed rate.
In Sec.~\ref{sec32}
we deduce a sufficient condition for the rate
to let the average of fidelity converge to $1$.
In Sec.~\ref{sec33}
we indicate that a small modification of the argument
in Secs.~\ref{sec31} and \ref{sec32} shows that
the same lower bound on the capacity is obtained from
the random coding of linear stabilizer codes.

\subsection{A definition of distance of a quantum channel from the
identity channel}\label{sec30}
We give a lower bound on the quantum capacity
in terms of the distance of a quantum channel from the identity channel.
Let us first review a definition of the distance of a quantum
channel from the identity channel
introduced in Ref.~\onlinecite{matsumotoerror},
then show some properties of the definition.
Let $\Gamma$ be a CP map on $\mathcal{L}(H_2)$.
\begin{definition}\label{def2}
Suppose that there exist a four-dimensional space $H_E$,
 $\ket{e_0} \in H_E$, and a unitary operator $U$
on $H_2 \otimes H_E$ such that
\begin{equation}
\Gamma(\rho) = \mathrm{Tr}_E [ U (\rho\otimes
\ket{e_0}\bra{e_0})U^*] \label{unirep2}
\end{equation}
for all $\rho \in \mathcal{S}(H_2)$.
Write $U$ as
\[
U = I \otimes L_I + \sigma_x \otimes L_x + \sigma_z \otimes L_z
+ \sigma_x\sigma_z \otimes L_{xz},
\]
where $L_I$, $L_x$, $L_z$, and $L_{xz}$ are linear operators
on $H_E$.
Then the distance $p(\Gamma)$ and $q(\Gamma)$
of the channel $\Gamma$ from the identity channel are
defined by
\begin{eqnarray*}
q(\Gamma) &=& \| L_I \ket{e_0}\|^2,\\
p(\Gamma) &=& \| L_x \ket{e_0}\|^2 +\| L_z \ket{e_0}\|^2 +
\| L_{xz} \ket{e_0}\|^2.
\end{eqnarray*}
\end{definition}

It is not clear whether the values of $p(\Gamma)$ and $q(\Gamma)$
are uniquely determined by $\Gamma$ alone,
that is, whether they are independent of choice of $U$ and
$\ket{e_0}$ in Eq.~(\ref{unirep2}).
In order to answer this question in Corollary \ref{pqunique},
we shall represent $p(\Gamma)$ and $q(\Gamma)$
using the operator-sum representation of $\Gamma$
induced by $U$ and $\ket{e_0}$.
\begin{proposition}\label{prop}
Let $\ket{e_0}$, \ldots, $\ket{e_3}$ be an orthonormal basis
of $H_E$, and
\[
A_i = \bra{e_i} U \ket{e_0}.
\]
By Eq.~(8.10) of Ref.~\onlinecite{chuangnielsen},
\[
\Gamma(\rho)= \sum_{i=0}^3 A_i \rho A_i^*
\]
for all $\rho \in \mathcal{S}(H_2)$.
Write $A_i$ as
\[
A_i = a_{i,I} I + a_{i,x}\sigma_x + a_{i,z}\sigma_z + a_{i,xz} \sigma_x\sigma_z.
\]
Then we have another representations of $p(\Gamma)$ and $q(\Gamma)$ as
\begin{eqnarray*}
p(\Gamma) &=& \sum_{i=0}^3 |a_{i,x}|^2 + |a_{i,z}|^2 + |a_{i,xz}|^2, \\
q(\Gamma) &=& \sum_{i=0}^3 |a_{i,I}|^2. 
\end{eqnarray*}
\end{proposition}
\noindent\emph{Proof.}
By definition of $A_i$,
\[
A_i = \langle e_i | L_I |e_0\rangle I + 
\langle e_i | L_x |e_0\rangle \sigma_x + 
\langle e_i | L_z |e_0\rangle \sigma_z + 
\langle e_i | L_{xz} |e_0\rangle \sigma_x\sigma_z.
\]
Therefore
\[
a_{i,I}=\langle e_i | L_I |e_0\rangle,\;
a_{i,x}=\langle e_i | L_x |e_0\rangle,\;
a_{i,z}=\langle e_i | L_z |e_0\rangle,\;
a_{i,xz}=\langle e_i | L_{xz} |e_0\rangle.
\]
Since $\ket{e_0}$, \ldots, $\ket{e_3}$ are an orthonormal
basis,
we have
\begin{eqnarray*}
q(\Gamma) &=& \| L_I \ket{e_0}\|^2\\
&=& \sum_{i=0}^3 |a_{i,I}|^2.
\end{eqnarray*}
The equality of $p(\Gamma)$ can be shown in a similar way.
\hfill$\blacksquare$

\begin{corollary}\label{pqunique}
The values of $p(\Gamma)$ and $q(\Gamma)$ do not depend
on choice of $U$ and $\ket{e_0}$ in Eq.~(\ref{unirep2}).
\end{corollary}
\noindent\emph{Proof.}
Let
\[
\Gamma(\rho) = \sum_{i=0}^3 B_i \rho B_i^*
\]
be another operator-sum representation of $\Gamma$.
By Theorem 8.2 of Ref.~\onlinecite{chuangnielsen},
there exists a $4\times 4$ unitary matrix $V$ such that
\[
\left(
\begin{array}{c}
B_0\\
\vdots\\
B_3\end{array}\right)
=
V
\left(
\begin{array}{c}
A_0\\
\vdots\\
A_3\end{array}\right).
\]
Write $B_i$ as
\[
B_i = b_{i,I} I + b_{i,x}\sigma_x + b_{i,z}\sigma_z + b_{i,xz} \sigma_x\sigma_z,
\]
and define $\vec{a}_I = (a_{0,I}$, \ldots, $a_{3,I})^T$,
$\vec{b}_I = (b_{0,I}$, \ldots, $b_{3,I})^T$.
Since $I$, $\sigma_x$, $\sigma_z$, and $\sigma_x\sigma_z$ are
linearly independent, we have $\vec{b}_I = V \vec{a}_I$.
Since $V$ is unitary, $\| \vec{b}_I\| = \| \vec{a}_I\|$,
which shows that $q(\Gamma)$ does not depend on choice of
representation.
The independence of $p(\Gamma)$ can be shown in a similar way.
\hfill$\blacksquare$

The following corollary drastically simplifies
the formula for the lower bound in Eq.~(\ref{condr2a}).
\begin{corollary}\label{pq1}
$p(\Gamma)+q(\Gamma) = 1$.
\end{corollary}
\noindent\emph{Proof.}
Notations as in Proposition \ref{prop}.
We have
\[
I = \sum_{i=0}^3 A_i^*A_i.
\]
Taking trace on the both side, we have
\begin{eqnarray*}
\mathrm{Tr}[I] & =&
\sum_{i=0}^3 \mathrm{Tr}[A_i^*A_i]\\
&=& \sum_{i=0}^3 \mathrm{Tr}[(a_{i,I} I + a_{i,x}\sigma_x + a_{i,z}\sigma_z + a_{i,xz} \sigma_x\sigma_z)^*\\
&&\mbox{} \times (a_{i,I} I + a_{i,x}\sigma_x + a_{i,z}\sigma_z + a_{i,xz} \sigma_x\sigma_z)]
\\
&=& \mathrm{Tr}[I] \sum_{i=0}^3 |a_{i,I}|^2 +|a_{i,x}|^2 + |a_{i,z}|^2 + |a_{i,xz}|^2
\\
&=& \mathrm{Tr}[I] (p(\Gamma)+q(\Gamma)). \qquad\blacksquare
\end{eqnarray*}

\subsection{Average of the fidelity over all the stabilizer codes}\label{sec31}
In this subsection we shall prove
that the average fidelity of the subcodes
of all $[[n$, $\lfloor Rn \rfloor]]$
stabilizer codes converges to $1$ as $n \rightarrow \infty$
under the conditions (\ref{condd}) and (\ref{condr}).
The proof is proceeded as follows:
\begin{enumerate}
\item For every stabilizer code,
there exists its subcode
whose fidelity of error correction is lower bounded by
Eq.~(\ref{preskillbound}).
The lower bound (\ref{preskillbound}) is expressed as
a sum  indexed by
error operators.
The average of the sum will be divided according to the weight of
error operators in Eq.~(\ref{eqc}).
\item It is easy to see
the part indexed by operators of larger weights
converges to $0$ as $n\rightarrow \infty$.
\item We shall show that the other part indexed by operators of smaller weights
converges to $0$ by the fact that
most of stabilizer codes can correct an error of small weight,
which will be rigorously proved in Eq.~(\ref{bnotmany}) from Lemma \ref{lemma4}.
\end{enumerate}

Let $\delta$ and $R$ be real numbers such that
\begin{eqnarray}
\lim_{n\rightarrow\infty}
\sum_{i=\lfloor  \delta n\rfloor+1}^{n}
{n\choose i} p(\Gamma)^i q(\Gamma)^{n-i}&=&0,
\label{condd}\\
1- \lim_{n\rightarrow \infty}\frac{\displaystyle\log_2 \left[
\sum_{i=1}^{\lfloor  \delta n\rfloor}
{n\choose i}p(\Gamma)^i q(\Gamma)^{n-i}
\sum_{j=0}^{i} {n\choose j} 3^j
\right]}{n} &>& R,\label{condr}
\end{eqnarray}
where $\lfloor x \rfloor$ denotes the largest integer $\leq x$.
\upshape

Let
\[
A_n = \{ C \subset \mathbf{F}_2^{2n} \mymid
C\mbox{ is linear},\;
\dim C = n - \lfloor Rn\rfloor,\;
C \subseteq C^\perp \}.
\]
Recall that we can construct an $[[n, \lfloor Rn\rfloor]]$
stabilizer code from every $C\in A_n$.
Note that $A_n$ is not empty because there exists
a self-orthogonal subspace of dimension $n$ in $\mathbf{F}_2^{2n}$.\cite{aschbacher00}
This subsection is devoted to show

\begin{proposition}\label{prop6}
If $R$ satisfies Eq.~(\ref{condr})
then there exists a sequence of subcodes of stabilizer codes
whose rates are greater than or equal to $R$ and
whose fidelity
converges to $1$ as $n\rightarrow\infty$.
\end{proposition}
Since the information rates of the subcode $Q'$ and $Q$
in Theorem \ref{thm2}
are asymptotically the same as $n\rightarrow \infty$,
it is suffice to show that the average of the fidelity
bound (\ref{preskillbound}) of $Q'$ over all the stabilizer codes
in $A_n$ converges to $1$ as $n\rightarrow \infty$.

Let $\ket{0_\mathrm{env}} = \ket{e_0}^{\otimes n}$, and
for $M = \sigma_{i_1} \otimes \cdots \otimes \sigma_{i_n}
\in \mathcal{E}$ let
\[
L_M = L_{i_1} \otimes \cdots \otimes L_{i_n},
\]
where $\sigma_I = I$ and
$\ket{e_0}$, $L_I$, $L_x$, $L_z$, and $L_{xz}$ are as defined
in Definition \ref{def2}.
For $C \in A_n$
we denote the set of uncorrectable errors of $C$ in $\mathcal{E}$
by $\mathcal{E}_\mathrm{unc}(C)$.
The average of the fidelity bound (\ref{preskillbound}) of $Q'$
over all the stabilizer codes in $A_n$ is
not less than
\begin{eqnarray}
&&\frac{1}{\sharp A_n}
\sum_{C\in A_n} \left(
1-
2\sum_{M \in \mathcal{E}_\mathrm{unc}(C)} \| L_M \ket{0_\mathrm{env}}\|^2\right)\nonumber\\
&=&1-\frac{2}{\sharp A_n}
\sum_{C\in A_n} \left(
\sum_{\renewcommand{\arraystretch}{0.1}\begin{array}{c}\scriptstyle M \in \mathcal{E}_\mathrm{unc}(C)\\\scriptstyle 1\leq w(M)\leq \lfloor \delta n\rfloor\end{array}}
\| L_M \ket{0_\mathrm{env}}\|^2+
\sum_{\renewcommand{\arraystretch}{0.1}\begin{array}{c}\scriptstyle  M \in \mathcal{E}_\mathrm{unc}(C)\\\scriptstyle w(M)> \lfloor \delta n\rfloor\end{array}} 
\|L_M \ket{0_\mathrm{env}}\|^2\right)\nonumber\\
&\geq&1-\left(\frac{2}{\sharp A_n}
\sum_{C\in A_n} 
\sum_{\renewcommand{\arraystretch}{0.1}\begin{array}{c}\scriptstyle M \in \mathcal{E}_\mathrm{unc}(C)\\\scriptstyle 1\leq w(M)\leq \lfloor \delta n\rfloor\end{array}}
\| L_M \ket{0_\mathrm{env}}\|^2\right) \nonumber\\*
&&\mbox{} -
2\sum_{\renewcommand{\arraystretch}{0.1}\begin{array}{c}\scriptstyle  M \in \mathcal{E}\\\scriptstyle w(M)> \lfloor \delta n\rfloor\end{array}}
\| L_M \ket{0_\mathrm{env}}\|^2. \label{eqc}
\end{eqnarray}
By the same argument as Ref.~\onlinecite{matsumotoerror},
one can show that
\[
\sum_{\renewcommand{\arraystretch}{0.1}\begin{array}{c}\scriptstyle  M \in \mathcal{E}\\\scriptstyle w(M)> \lfloor \delta n\rfloor\end{array}}
\| L_M \ket{0_\mathrm{env}}\| \leq
\sum_{i=\lfloor  \delta n\rfloor+1}^{n}
{n\choose i} p(\Gamma)^i q(\Gamma)^{n-i},
\]
which converges to $0$ as $n\rightarrow\infty$
by the condition (\ref{condd}).

We shall calculate an upper bound for the second term in
Eq.~(\ref{eqc}).
For $M \in \mathcal{E}$ we define
\[
B_n(M) = \{ C \in A_n \mymid M \in \mathcal{E}_\mathrm{unc}(C) \}.
\]
It follows that
\begin{eqnarray}
&& \frac{1}{\sharp A_n}
\sum_{C\in A_n} 
\sum_{\renewcommand{\arraystretch}{0.1}\begin{array}{c}\scriptstyle M \in \mathcal{E}_\mathrm{unc}(C)\\\scriptstyle 1\leq w(M)\leq \lfloor \delta n\rfloor\end{array}}
\| L_M \ket{0_\mathrm{env}}\|^2\nonumber\\
&\leq&
\frac{1}{\sharp A_n}
\sum_{\renewcommand{\arraystretch}{0.1}\begin{array}{c}\scriptstyle M \in \mathcal{E}\\\scriptstyle 1\leq w(M)\leq \lfloor \delta n\rfloor\end{array}}
\sharp B_n(M) \| L_M \ket{0_\mathrm{env}}\|^2. \label{eqa}
\end{eqnarray}
Note that we omitted the factor $2$ from Eq.~(\ref{eqc})
for simplicity, because we shall show that
the right hand side of Eq.~(\ref{eqa}) converges to $0$
and factor $2$ is negligible.

We shall give an upper bound for $\sharp B_n(M)$.
To estimate $\sharp B_n(M)$
we shall introduce Lemma \ref{lemma4}.
In the proof of Lemma \ref{lemma4} we use the Witt theorem,
so we review it.

\begin{theorem}[Witt]
Let $K$ be a field, $V_1$, $V_2$ finite-dimensional
$K$-linear spaces, and $\tau_1$, $\tau_2$
symplectic forms on $V_1$, $V_2$, respectively.
An injective linear map $T : V_1 \rightarrow V_2$ is said to be
an \emph{isometry} if
\[
\tau_1(x,y) = \tau_2(Tx, Ty).
\]
Let $W_1$ be a subspace of $V_1$.
If there exists a bijective isometry from $V_1$ to $V_2$
and an isometry $T_{W_1} : W_1 \rightarrow V_2$,
then there exists an isometry $T_{V_1} : V_1 \rightarrow V_2$
such that the restriction of $T_{V_1}$ to $W_1$ is
equal to $T_{W_1}$.
The same result also holds when $\tau_1$, $\tau_2$ are
hermitian forms.
\end{theorem}

\noindent\emph{Proof.} See Sec.~20 of Ref.~\onlinecite{aschbacher00}.
\hfill$\blacksquare$

\begin{lemma}\label{lemma4}
For $M \in E\setminus\{\pm I\}$,
let $A_n(M) = \{C \in A_n \mymid
f(M) \in C^\perp\setminus C\}$.
We have
\[
\sharp A_n(M) \leq
(1/2^{n-\lfloor Rn\rfloor})
\frac{1-2^{-2\lfloor Rn\rfloor}}{1-2^{-2n}}\sharp A_n
<
\sharp A_n / 2^{n-\lfloor Rn\rfloor}.
\]
\end{lemma}

\noindent\emph{Proof.}
Let $\mathrm{Sp}_n(\mathbf{F}_2)$
be the group of bijective linear maps on $\mathbf{F}_2^{2n}$
preserving the symplectic form (\ref{symplecticform}).
For every pair of spaces $C_1$, $C_2 \in A_n$,
every bijective linear map from $C_1$ to $C_2$ is an
isometry.
Consequently,
there exists $\sigma \in 
\mathrm{Sp}_n(\mathbf{F}_2)$ such that
$\sigma C_1 = C_2$ by the Witt theorem.
A similar argument shows that
there exists $\sigma' \in \mathrm{Sp}_n(\mathbf{F}_2)$ such that
$\sigma' (\vec{a}|\vec{b}) = (\vec{a}'|\vec{b}')$
for every pair of nonzero vectors $(\vec{a}|\vec{b})$,
$(\vec{a}'|\vec{b}')\in \mathbf{F}_2^{2n}$.

It follows that
\begin{eqnarray*}
&&\sharp A_n(M) \\
&=& \sharp \{C \in A_n \mymid
f(M) \in C^\perp\setminus C\}\\
&=& \sharp \{ \alpha C_1 \mymid
f(M) \in (\alpha C_1)^\perp\setminus\alpha C_1, \alpha\in \mathrm{Sp}_n(\mathbf{F}_2)\}\\
&=& \sharp \{\alpha C_1 \mymid
\beta(f(M)) \in (\alpha C_1)^\perp\setminus\alpha C_1, \alpha\in \mathrm{Sp}_n(\mathbf{F}_2)\},
\end{eqnarray*}
where $C_1$ (resp.\ $\beta$) is an arbitrary fixed
element in $A_n$ (resp.\ $\mathrm{Sp}_n(\mathbf{F}_2)$).
Therefore
$\sharp A_n(M)$ is the same among every
nonzero $f(M)$.

Since $\sharp (C^\perp\setminus C) = 2^{n+\lfloor Rn\rfloor}-
2^{n-\lfloor Rn\rfloor}$,
there are $(2^{n+\lfloor Rn\rfloor}-
2^{n-\lfloor Rn\rfloor}) \sharp A_n$ pairs of
$((\vec{a}|\vec{b}),C)$ such that $(\vec{a}|\vec{b}) \in C^\perp
\setminus C$ and $C\in A_n$.
Thus if $M\neq \pm I$ then 
\begin{eqnarray*}
\sharp A_n(M)&\leq&
\frac{2^{n+\lfloor Rn\rfloor}-
2^{n-\lfloor Rn\rfloor}}{2^{2n}-1}\sharp A_n \\
&=&
(1/2^{n-\lfloor Rn\rfloor})
\frac{1-2^{-2\lfloor Rn\rfloor}}{1-2^{-2n}}\sharp A_n\\
&<&
\sharp A_n / 2^{n-\lfloor Rn\rfloor}. \qquad{\blacksquare}
\end{eqnarray*}

\begin{remark}\label{remark5} From Lemma \ref{lemma4}
we can improve the quantum Gilbert-Varshamov bound
slightly.
There exists an $[[n,k,d]]$ stabilizer code if
\begin{equation}
\frac{1-2^{-2k}}{1-2^{-2n}} \cdot
\frac{1}{2^{n-k}}
\sum_{i=1}^{d-1}3^i 
{n\choose i} < 1. \label{qgv}
\end{equation}
The proof is as follows:
For each error $M \in \mathcal{E}$,
$A_n(M)$ is equal to
the set of stabilizer codes unable to detect $M$ as an error.
Therefore,
by replacing $\lfloor Rn\rfloor$ with $k$ in Lemma \ref{lemma4},
we see that if Eq.~(\ref{qgv}) holds then there is at least one
stabilizer code $C$ able to detect all the errors $M$ with $w(M) <d$,
which means that the minimum distance of $C$ is at least $d$.
The idea behind this proof
already appeared in
the original proof of the quantum Gilbert-Varshamov bound
for stabilizer codes.\cite{calderbank97}
Observe that our bound is slightly better than
the quantum Gilbert-Varshamov bound
for general codes,\cite{ekert96}
which implies that an $[[n,k,d]]$ quantum code exists if
\[
\frac{1}{2^{n-k}} \sum_{i=0}^{d-1} 3^i {n\choose i}<1.
\]
\end{remark}

By Eq.~(\ref{uncorrectableset}),
$M \in \mathcal{E}$ belongs to $\mathcal{E}_\mathrm{unc}(C)$
only if there exists $M' \in \mathcal{E}$ such that
$w(M') \leq w(M)$, $Mf^{-1}(C^\perp) = M'f^{-1}(C^\perp)$,
and $M f^{-1}(C) \neq M' f^{-1}(C)$.
A space $C \in A_n$ belongs to $B_n(M)$ only
if there exists $M'\in \mathcal{E}$ such that
$w(M') \leq w(M)$ and 
$M^{-1} M' \in f^{-1}(C^\perp\setminus C)$.
The last condition is equivalent to
$C \in A_n(M^{-1}M')$.
Since there are
\[
\sum_{j=0}^{w(M)} {n\choose j}
3^j
\]
operators $M' \in \mathcal{E}$ such that $w(M') \leq w(M)$,
it follows that
\begin{eqnarray}
\sharp B_n(M) &\leq&
\sum_{\renewcommand{\arraystretch}{0.1}\begin{array}{c}\scriptstyle  M' \in \mathcal{E}\\\scriptstyle w(M')\leq w(M)\end{array}}
\sharp A_n(M^{-1}M')\nonumber\\
&\leq&
\sum_{\renewcommand{\arraystretch}{0.1}\begin{array}{c}\scriptstyle  M' \in \mathcal{E}\\\scriptstyle w(M')\leq w(M)\end{array}}
\frac{\sharp A_n}{2^{n-\lfloor Rn\rfloor}}\nonumber\\
&\leq&
\frac{\sharp A_n}{2^{n-\lfloor Rn\rfloor}}
\sum_{j=0}^{w(M)} \left(\begin{array}{c}n\\ j\end{array}\right)
3^j.\label{bnotmany}
\end{eqnarray}

An upper bound for Eq.~(\ref{eqa}) is derived as follows:
\begin{eqnarray}
&&\frac{1}{\sharp A_n}
\sum_{\renewcommand{\arraystretch}{0.1}\begin{array}{c}\scriptstyle M \in \mathcal{E}\\\scriptstyle 1\leq w(M)\leq \lfloor \delta n\rfloor\end{array}}
\sharp B_n(M) \| L_M \ket{0_\mathrm{env}}\|^2\nonumber\\
&\leq&
\frac{1}{\sharp A_n} \sum_{\renewcommand{\arraystretch}{0.1}\begin{array}{c}\scriptstyle M \in \mathcal{E}\\\scriptstyle 1\leq w(M)\leq \lfloor \delta n\rfloor\end{array}}
\frac{\sharp A_n}{2^{n-\lfloor Rn\rfloor}}
\sum_{j=0}^{w(M)} {n\choose j}
3^j \| L_M \ket{0_\mathrm{env}}\|^2\nonumber\\
&=&
\frac{1}{2^{n-\lfloor Rn\rfloor}}
\sum_{\renewcommand{\arraystretch}{0.1}\begin{array}{c}\scriptstyle M \in \mathcal{E}\\\scriptstyle 1\leq w(M)\leq \lfloor \delta n\rfloor\end{array}}
\sum_{j=0}^{w(M)} {n\choose j}
3^j \| L_M \ket{0_\mathrm{env}}\|^2
\label{eqb}.
\end{eqnarray}
For an integer $0\leq i \leq n$,
by the same argument as,\cite{matsumotoerror}
one can show that
\[
\sum_{\renewcommand{\arraystretch}{0.1}\begin{array}{c}\scriptstyle 
M \in \mathcal{E}\\\scriptstyle w(M)=i\end{array}} \| L_M \ket{0_\mathrm{env}}
\|^2
= {n\choose i}
p(\Gamma)^i q(\Gamma)^{n-i}.
\]
Therefore Eq.~(\ref{eqb}) is equal to
\[
\frac{1}{2^{n-\lfloor Rn\rfloor}}
\sum_{i=1}^{\lfloor \delta n\rfloor}
{n\choose i}
p(\Gamma)^i q(\Gamma)^{n-i}
 \sum_{j=0}^{i} {n\choose j}
3^j,
\]
which converges to $0$ as $n\rightarrow\infty$
by the condition (\ref{condr}).

\subsection{Achievable rate by general stabilizer codes}\label{sec32}
In the previous subsection,
we have shown that
if the rate $R$ satisfies Eq.~(\ref{condr})
then there exists at least one sequence of subcodes of stabilizer codes
of the rate $R$ such that the average of fidelity converges
to $1$.
In this subsection we shall simplify
Eqs.~(\ref{condd}) and (\ref{condr}) with which
we can easily compute a lower bound on the capacity
of the channel $\Gamma$.

We shall deduce a sufficient condition for $\delta$
to satisfy Eq.~(\ref{condd}).
By Appendix A of Ref.~\onlinecite{peterson}, for
$0\leq \epsilon < \lambda \leq 1$ we have
\[
\sum_{i=\lambda n}^n {n\choose i}
\epsilon^i (1-\epsilon)^{n-i} \leq
2^{-nD(\lambda \| \epsilon)},
\]
where $D(\lambda \| \epsilon)$ is the classical relative entropy
defined by
\[
\lambda \log_2\frac{\lambda}{\epsilon}
+(1-\lambda)\log_2\frac{1-\lambda}{1-\epsilon}.
\]
Since $p(\Gamma)+q(\Gamma)=1$ by
Corollary \ref{pq1},
the condition (\ref{condd}) holds if
\begin{equation}
\delta > p(\Gamma).\label{condd2}
\end{equation}

The term
inside of $\log_2$ in Eq.~(\ref{condr})
can be bounded as follows:
\begin{eqnarray*}
&& \sum_{i=1}^{\lfloor \delta n\rfloor}{n \choose i}p(\Gamma)^i q(\Gamma)^{n-i}
 \sum_{j=0}^{i}{n \choose j}3^j \\
&\leq& \sum_{i=0}^{n}{n \choose i}p(\Gamma)^i q(\Gamma)^{n-i}
 \sum_{j=0}^{\lfloor \delta n\rfloor}{n \choose j}3^j\\
&=& [p(\Gamma)+q(\Gamma)]^n
 \sum_{j=0}^{\lfloor \delta n\rfloor}{n \choose j}3^j\\
&=& \sum_{j=0}^{\lfloor \delta n\rfloor}{n \choose j}3^j
\mbox{ (by Corollary \ref{pq1})} \\
&\leq& (\delta n+1) {n \choose \lfloor \delta n\rfloor}3^{\delta n} \\
&\leq& (\delta n+1) \exp_2[n H_\mathrm{e}(\delta)] 3^{\delta n}
\mbox{ (by Appendix B of Ref.~\onlinecite{peterson})}\\
&=& (\delta n + 1)\exp_2\{n[H_\mathrm{e}(\delta)+\delta\log_2 3]\},
\end{eqnarray*}
where $H_\mathrm{e}$ is the binary entropy function.

{From} Proposition \ref{prop6}, Eq.~(\ref{condd2}), and the observations above,
we see that the capacity of the channel $\Gamma$ is at least
\begin{equation}
1 - \{H_\mathrm{e}[p(\Gamma)]+ p(\Gamma)\log_2 3\}.
\label{condr2a}
\end{equation}
Note that the same lower bound on the capacity
can also be obtained by the method of Bennett et~al.,\cite{bennett96}
though they stated their result only for the depolarizing channel.
However, Bennett et~al.\cite{bennett96} did not
address the achievability of the bound (\ref{condr2a})
with stabilizer codes,
which is the main focus of this paper.

We shall compare our bound on the capacity [Eq.~(\ref{condr2a})] with the conventional bound for a general
memoryless channel
derived from the quantum Gilbert-Varshamov bound\cite{calderbank97,ekert96}
and the fidelity bounds for $t$-error-correcting codes.\cite{knill97,matsumotoerror}
Suppose that we have a sequence of $\lfloor \delta n_i\rfloor$-error-correcting
quantum codes of length $n_i$ with $\lim_{i\rightarrow\infty}
n_i = \infty$.
The condition (\ref{condd2}) is sufficient in order that
the fidelity of error correction
by $\lfloor \delta n_i\rfloor$-error-correcting codes
converges to $1$ as $i\rightarrow\infty$.
By the quantum Gilbert-Varshamov bound the derived lower bound
on the capacity is
\[
1 -\{ H_\mathrm{e}[2p(\Gamma)] + 2p(\Gamma) \log_2 3\},
\]
which is always smaller than Eq.~(\ref{condr2a}).

When the channel $\Gamma$ is the depolarizing channel
of the fidelity parameter $f$,
$p(\Gamma)=1-f$ and $q(\Gamma)=f$.
The proposed lower bound [Eq.~(\ref{condr2a})]
for the capacity
is
\[
1- [H_\mathrm{e}(1-f)+(1-f) \log_2 3],
\]
which coincides with the lower bound given in Ref.~\onlinecite{bennett96}.
It is not clear to the authors whether
our lower bound can be improved by the method in
Ref.~\onlinecite{divincenzo98}.

Our analysis for the quantum capacity can be
generalized to the capacity of an $l$-adic channel
using the $l$-adic stabilizer codes\cite{knill96a,rains97}
in a straightforward manner when $l$ is prime.
The quantum Gilbert-Varshamov bound for $l$-adic
stabilizer codes
can also be proved by Lemma \ref{lemma4}.

\subsection{Achievable rate by linear stabilizer codes}\label{sec33}
In this subsection we shall show that
the achievable rate (\ref{condr2a})
by subcodes of general stabilizer codes
can also be achieved by those of linear stabilizer codes,
which shows the asymptotic optimality of linear stabilizer codes
among general ones.
As a byproduct we establish an analogue of Gilbert-Varshamov bound
for linear stabilizer codes.

Let
\[
A'_n = \{ C \subset \mathbf{F}_4^n \mymid
C\mbox{ is $\mathbf{F}_4$-linear},\;
\dim_{\mathbf{F}_4} C = \lfloor (n-Rn)/2 \rfloor,\;
C \subseteq C^\perp\}.
\]
Recall that we can construct an $[[n, n-2\lfloor (n-Rn)/2 \rfloor]]$
linear stabilizer code from every $C \in A'_n$.
Note that $A'_n$ is not empty because there exists
a self-orthogonal subspace of dimension $\lfloor n/2\rfloor$
in $\mathbf{F}_4^n$ (see Proposition 2.3.2 in Ref.~\onlinecite{kleidman90}).
For $M \in \mathcal{E}$, define
\begin{eqnarray*}
A'_n(M) &=& \{ C \in A'_n \mymid g(M) \in C^\perp \setminus C \},\\
B'_n(M) &=& \{ C \in A'_n \mymid M \mbox{ is uncorrectable by }C\}.
\end{eqnarray*}
By these definitions of $A'_n(M)$ and $B'_n(M)$,
all the argument except Lemma \ref{lemma4} in the previous subsections
can be used for showing that the rate (\ref{condr2a})
is achieved by subcodes of linear stabilizer codes.
In this subsection we prove an upper bound (Lemma \ref{lemma8}) for
$\sharp A'_n(M)$ that can be used as a substitute for Lemma \ref{lemma4}.

\begin{lemma}\label{lemma8}
Define $\tau$ by Eq.~(\ref{hermitianinner}).
The number of nonzero vectors $\vec{x} \in \mathbf{F}_4^n$
such that $\tau(\vec{x}$, $\vec{x}) = 0$
is $2^{2n-1}+(-1)^n 2^{n-1}-1$.
\end{lemma}

\noindent\emph{Proof.} See the
proof of Proposition 2.3.3 in Ref.~\onlinecite{kleidman90}.
\hfill$\blacksquare$

\begin{lemma}
Let $u = \lfloor (n-Rn)/2 \rfloor$.
For $M \in E \setminus \{\pm I\}$
\begin{eqnarray*}
\sharp A'_n(M) &\leq&
\frac{4^{n-u}-4^u}{\min\{2^{2n-1}+(-1)^n 2^{n-1}-1,
2^{2n-1} - (-1)^n 2^{n-1}\}} \sharp A'_n\\
&\leq& \frac{4^{n-u}-4^u}{2^{2n-1}- 2^{n-1}-1}\sharp A'_n.
\end{eqnarray*}
\end{lemma}

\noindent\emph{Proof.}
Let $\mathrm{GU}_n(\mathbf{F}_4)$ be the group
of bijective linear maps on $\mathbf{F}_4^n$
that preserve the value
of the hermitian form $\tau$.
For every pair of spaces $C_1, C_2 \in A'_n$,
every bijective linear map from $C_1$ to $C_2$ is
an isometry.
Thus there exists $\sigma \in \mathrm{GU}_n(\mathbf{F}_4)$
such that $\sigma C_1 = C_2$ by the Witt theorem.
For a pair of nonzero vectors $\vec{x}$, $\vec{y}
\in \mathbf{F}_4^n$ with $\tau(\vec{x}$, $\vec{x}) =
\tau(\vec{y}$, $\vec{y}) = 0$,
a similar argument shows
that there exists $\sigma \in \mathrm{GU}_n(\mathbf{F}_4)$
such that $\sigma \vec{x} = \vec{y}$.

We want to show that
for a pair of vectors 
$\vec{x}$, $\vec{y}
\in \mathbf{F}_4^n$ with $\tau(\vec{x}$, $\vec{x}) \neq 0$
and
$\tau(\vec{y}$, $\vec{y}) \neq 0$,
there exists $\sigma \in \mathrm{GU}_n(\mathbf{F}_4)$
such that $\sigma \vec{x} = \vec{y}$.
Since $\tau$ is a hermitian form,
$\tau(\vec{x}$, $\vec{x}) \in \mathbf{F}_2$.
Therefore $\tau(\vec{x}$, $\vec{x}) = \tau(\vec{y}$, $\vec{y})
=1$, and
there exists $\sigma \in \mathrm{GU}_n(\mathbf{F}_4)$
such that $\sigma \vec{x} = \vec{y}$ by the 
Witt theorem.

A similar argument to the proof of Lemma \ref{lemma4} shows that
for $M \in E \setminus \{\pm I\}$ we have
\begin{eqnarray*}
\sharp A'_n(M) &\leq& \frac{4^{n-u}-4^u}{2^{2n-1}+ (-1)^n 2^{n-1}-1}
 \sharp A'_n \qquad\mbox{if }\tau(g(M),g(M)) = 0,
\\
\sharp A'_n(M) &\leq& 
\frac{4^{n-u}-4^u}{2^{2n-1} -(-1)^n 2^{n-1}} \sharp A'_n
\qquad\mbox{if }\tau(g(M),g(M)) \neq 0. \qquad{\blacksquare}
\end{eqnarray*}

\begin{remark}\label{gvlinear} From
Lemma \ref{lemma8} we can show that there exists an $[[n,k,d]]$ linear stabilizer code if
$k$ is even and
\[
\frac{2(1-2^{-2k})}{1-2^{-n}-2^{-2n+1}} \cdot
\frac{1}{2^{n-k}}
\sum_{i=1}^{d-1}3^i 
{n\choose i} < 1,
\]
which is asymptotically the same as
the quantum Gilbert-Varshamov bound
for general quantum codes.\cite{ekert96}
\end{remark}

\begin{remark}\label{remark9}
The connection between stabilizer codes and classical codes
over $\mathbf{F}_4$ was generalized to nonbinary case
in Refs.~\onlinecite{ashikhmin00,bierbrauer00,matsumotouematsu00}.
The argument in this subsection can be extended to
linear $l$-adic stabilizer codes for a prime $l$
with the following exception:
In the proof of Lemma \ref{lemma8},
there does not always exist $\sigma \in \mathrm{GU}_n(\mathbf{F}_{l^2})$
such that $\sigma\vec{x}=\vec{y}$ for
a pair of vectors $\vec{x}$, $\vec{y} \in \mathbf{F}_{l^2}^n$
with $\tau(\vec{x}$,
$\vec{x}) \neq 0$ and $\tau(\vec{y}$,
$\vec{y}) \neq 0$.
However,
there always exists $\sigma \in \mathcal{U}$
such that $\sigma\vec{x}=\vec{y}$,
where $\mathcal{U}$ is the group
generated by $\mathrm{GU}_n(\mathbf{F}_{l^2})$
and nonzero scalar multiples of the identity map on $\mathbf{F}_{l^2}^n$.
\end{remark}

\section*{Acknowledgment}
We would like to thank Dr.\ Mitsuru Hamada
for drawing our attention to the random coding of stabilizer codes
and for providing detailed comments on this paper,
and Dr.\ Masahito Hayashi, Dr.\ Keiji Matsumoto, and
Dr.\ Tomohiro Ogawa for helpful discussions.


\end{document}